 [2003/12/12 5.2/AAS markup document class]%
\documentclass[12pt,preprint]{aastex}
\usepackage{natbib}
\usepackage{subfig}

\shorttitle{On the obscuration of the SMBH}
\shortauthors{Hueyotl-Zahuantitla et al.}

\begin{document}

\title{ On the fate of the matter reinserted within young nuclear stellar clusters}

%\author{Filiberto Hueyotl-Zahuantitla\altaffilmark{1}, Jan Palou\v s\altaffilmark{1}, Richard W\"unsch\altaffilmark{1} }
\author{Filiberto Hueyotl-Zahuantitla, Jan Palou\v s, Richard W\"unsch}
\affil{Astronomical Institute, Academy of Sciences of the Czech Republic, Bo\v cn\'i II 1401, 141 31 Prague, Czech Republic.}

\and

\author{Guillermo Tenorio-Tagle\altaffilmark{2} and Sergiy Silich\altaffilmark{2}}
\affil{Instituto Nacional de Astrof\'isica Optica y Electr\'onica, AP 51, 72000 Puebla, M\'exico}
\email{filibert@asu.cas.cz}

%\altaffiltext{1}{Astronomical Institute, Academy of Sciences of the Czech Republic, Bo\v cn\'i II 1401, 141 31 Prague, Czech Republic; filibert@asu.cas.cz }
%\altaffiltext{2}{Instituto Nacional de Astrof\'isica Optica y Electr\'onica, AP 51, 72000 Puebla, M\'exico; gtt@inaoep.mx}

%\myemail

\begin{abstract}
This paper presents a hydrodynamical model describing the evolution of the gas reinserted by stars within a rotating young nuclear star cluster (NSC). We explicitly consider the impact of the stellar component to the flow by means of a uniform  insertion of mass and energy within the stellar cluster.  The model includes the gravity force of the stellar component and a central supermassive black hole (SMBH), and accounts for  the heating from the central source of radiation and the radiative cooling of the thermalized gas.  By using a set of parameters typical for NSCs and SMBHs in Seyfert galaxies our simulations show that a filamentary/clumpy  structure is formed in the inner part of the cluster.  This ``torus"  is  Compton thick and covers a large fraction of the sky (as seen from the SMBH).  In the outer parts of the cluster  a powerful wind is produced,  that  inhibits the infall of matter from  larger scales and thus  the NSC-SMBH interplay  occurs in isolation.

\end{abstract}


\section{Introduction}\label{s1}
The origin of the obscuring matter in active galactic nuclei (AGN) is one of the main challenges in modern astrophysics. The Unified scheme for AGNs requires  of a dusty torus  to explain the two main types: non-obscured  (type 1) and obscured (type 2) AGN  \citep{1993ARA&A..31..473A,1995PASP..107..803U}. The model claims that the  highly energetic central activity of AGNs is powered by mass accretion onto a supermassive black hole (SMBH) with mass $M_{\rm{SMBH}}=10^{6}-10^{10}~M_{\odot}$, and that the distinction between different types of AGNs is simply determined by the orientation of the torus.
The origin, structure, and dynamics of the torus remains as one of the key unsolved problems in AGN physics, which presumably is also related to the SMBH feeding and feedback. Some models assume uniform gas distribution for the obscuring torus, whose thickness is supported by IR radiation pressure \citep[e.g.][and references therein]{1992ApJ...401...99P,2007ApJ...661...52K}. Other models assume that the circumnuclear obscuring medium is clumpy, see for example \citet{1988ApJ...329..702K} and  \citet{2002ApJ...570L...9N}.  However, the mechanism  supporting the vertical thickness (in homogeneous models) and  motions of the clumps (in clumpy models) is still under debate.
\citet{2006ApJ...648L.101E} suggested that the torus is simply an outflow of dusty and optically thick clumps coming from the accretion disc  . \citet{1994ApJ...434..446K} proposed a model where a magneto-centrifugal wind is responsible for the obscuration.   \citet{2011ApJ...741...29D} proposed a model in which the obscuration is produced by a dusty wind driven by infrared radiation pressure  from a dense torus. The torus as a wind model does not suffer from the vertical structure problem, but the origin of the wind is still unclear.  \citet{2012MNRAS.419.1238N} suggest that the obscuring matter comes from fragmentation of  solid bodies (asteroids, comets and terrestrial-like planets) in the vicinity of the SMBH.  \citet{2012ApJ...758...66W},  based on three dimensional simulations including radiative feedback from the AGN, describe the formation of a turbulent  torus from the interaction of back-flows in a bipolar  fountain, starting from a preexisting rotationally supported thin disk.   \citet{2012arXiv1206.4059L} provided numerical simulations for the accretion flows with angular momentum which is sufficient to inhibit the accretion if the viscous processes are negligible, and to form a torus from the centrifugal gas.   Most of the  above models are dedicated mainly to explain the mechanism for  vertical support of the torus and they are based on the assumption that a cold gaseous structure already exists.  They also neglect the effects that the stellar feedback may provide to the inflow onto the SMBH. 

\citet{2002ApJ...566L..21W} and \citet{2009MNRAS.393..759S} suggested that clumpy tori form with  gas reinserted by stars within massive nuclear clusters  during the late (post-supernovae) evolution. This model was motivated by the fact that stellar activity in the vicinity of the central SMBH has been found in a variety of Hubble type galaxies \citep[see, for example,][and references therein] {2003ApJ...588L..13F,2008AJ....135..747G,2008ApJ...678..116S,2009ApJ...691L.142K}.
Such nuclear starbursts are compact ($< 50$~pc) and have masses in the range $10^6 - 10^9~M_{\odot}$ \citep{2007ApJ...671.1388D,2008ApJ...677..895W,2010ApJ...714..713S}. Some of them show complicated star formation histories \citep{2006ApJ...649..692W} and present evidence for global rotation, up to 45 km s$^{-1}$, and stellar velocity dispersion of up to 120~km s$^{-1}$. The \citet{2009MNRAS.393..759S} model seems to be in agreement with \citet{2007ApJ...671.1388D} and \citet{2010MNRAS.405..933W} who claim that the accretion rates and thus the AGN luminosities rise rapidly at the late stage of the nuclear starburst evolution. However, this model does not explain the coincidence of luminous AGNs with young (ages less than 40~Myr) nuclear starbursts, as it is the case of NGC 1097 (see Figure 11 in Davies et al. 2007).  
Here we show that massive and compact young NSCs with a central  SMBH can form filamentary/clumpy gaseous tori, and that the  sizes of such tori are in the range inferred by near-infrared observations of Seyfert galaxies \citep{2004Natur.429...47J,2005AJ....130.1472P} and with the sizes derived from their spectral energy distributions \citep[SEDs,][and references therein]{2011ApJ...736...82A}. These tori are consistent also with the compact sizes of the Compton thick medium estimated from X-ray observations of Seyfert 2 galaxies, type 2 AGNs \citep{1999ApJ...522..157R}.

The paper is organized as follows. Section 2 deals with all the ingredients of the hydrodynamical model. Details regarding the numerical scheme are given in section 3.  Section 4 describes the hydrodynamic solution which leads to the formation of the torus. We start the discussion with a model without the central radiation field. Then we include the central source of radiation and calculate column densities and possible obscuration fraction  of the torus. We also compare analytic estimates of the centrifugal barrier with the numerical results. Section 5 discusses the impact of the NSC wind on the host galaxy and section 6 presents our  conclusions.

\section{The physical model}\label{s2}
We consider the  accretion flow onto a SMBH embedded into a young stellar cluster which rotates like a solid body.  The key point for our model is that the matter reinserted within young, massive, and compact star clusters could evolve in a catastrophic cooling regime which is very different from the \citet{1985Natur.317...44C} adiabatic solution \citep[see][hereafter S04, TT07 and W08, respectively]{2004ApJ...610..226S,2007ApJ...658.1196T,2008ApJ...683..683W}. In the Chevalier and Clegg solution the  thermalization of the matter ejected by massive stars inside the cluster leads to a  high central pressure with an outward pressure gradient  that steadily accelerates the gas from
zero velocity at the center (i.e., the stagnation point is at the center) to the sound speed at the cluster edge.  The reinserted matter exits the cluster as a free wind approaching its terminal velocity, which is twice the sound speed at the cluster border.  If radiative cooling is considered, the gas in the central zones of massive clusters cools down and eventually becomes thermally unstable. This is because the average density of the gas increases linearly with the cluster mass while the cooling rate inside the cluster volume grows as a square function of the stellar cluster mass. Consequently, the central pressure drops and the gas cannot be accelerated outwards. The stagnation point moves out of the cluster center and the solution becomes bimodal: within the stagnation volume the thermal instability leads to mass accumulation, while in the outer parts of the cluster 
a stationary cluster wind is established. The size of the stagnation zone becomes larger as one considers more massive clusters where strong radiative energy losses  favor  
the frequent generation of cold parcels  of gas (see TT07 for the semi-analytic procedure for calculating the stagnation radius). This leads to mass accumulation  and eventually to star formation and thus to a positive star formation feedback \citep[see][]{2005ApJ...628L..13T}.

 In the presence of a central SMBH the stagnation point is always out from the center \citep[][here after S08 and HZ10, respectively]{2008ApJ...686..172S, 2010ApJ...716..324H} and thus the flow in the vicinity of the central SMBH is always bimodal.  Matter inserted within the stagnation volume forms an accretion flow whereas the mass inserted between the stagnation radius and the cluster edge drives the cluster wind. In such cases, the stagnation radius, $R_{\rm{st}}$, is defined by the balance between the outward pressure gradient (strongly affected by radiative cooling) and the gravity force  that makes the accretion flow very different from the classic Bondi solution.  All matter reinserted within the stagnation zone remains bound to the cluster and thus defines the upper limit to the mass accretion rate onto the SMBH.
In thermally unstable bimodal cases, as radiative cooling becomes more important, the stagnation point moves to a larger radius increasing substantially the mass accretion rate. Here by using typical values for the masses and sizes of NSCs, and masses of the SMBH in Seyfert galaxies, we explore the formation of the torus from the mass inserted within a rotating young NSCs with a central SMBH.

Our physical model consists of a spherically-symmetric young NSC of radius $R_{\rm{NSC}}$ and mass $M_{\rm{NSC}}$ with a homogeneous distribution of stars and with a central SMBH of mass $M_{\rm{SMBH}}$. It accounts for the gravity pull from the stellar component and from the black hole. We consider a constant injection of mechanical energy ($L_{\rm{NSC}}$) and mass ($\dot M_{\rm{NSC}}$) within the cluster volume via SNe II and stellar winds.  
Matter is inserted within the star cluster with a finite angular momentum given as the solid-body rotation around the polar axis: ${\bf v}_{\rm{rot}}=\omega r \sin{\theta}\hat \phi$, where $\omega$ is the angular frequency, $r$ and $\theta$ are the radial and polar coordinates, respectively, and $\hat \phi$ is the unit vector in the direction $\phi$. We assume that the rotation velocity of the star cluster is small compared to the dispersion velocity of individual stars and thus we disregard the star cluster flattening due to its rotation.  Radiative cooling is one of the main ingredients of the model and it is considered in all the computational domain. 

One of the main features of AGNs is their strong emission of ionizing radiation. A proper treatment of the emission from the central source requires to know the intensity and the spectral energy distribution of the radiation field, and how it propagates through the ambient medium. 
 Here we consider only the X-ray luminosity $L_{X}$, since such photons can penetrate deeper into high density regions shielded from UV photons.  We consider a constant Eddington ratio $L_{X}/L_{\rm Edd}=0.08$ during the calculations ($L_{\rm Edd}\sim 1.3\times 10^{38}$ erg s$^{-1}$). This value is in the range 0.01--0.1 used by \citet{2012ApJ...758...66W}. In this way, the model includes Compton and X-ray heating from the central source and  the radial component of the acceleration due to radiation pressure. These quantities were explicitly calculated using the ray-tracing method described in Appendix \ref{A}. 
 
 The model implicitly accounts for shock heating by assuming full thermalization of stellar winds and SNe kinetic energy due to random interactions of the ejecta from massive stars that leads to gas temperatures of a few $10^{7}$ K. The reinserted gas at such temperature and relatively high density is not in thermal equilibrium.
  
 In all calculations we assume a maximum rotation velocity along the equator $v_{\rm rot=}$ 50 km s$^{-1}$ at the star cluster edge. The energy and mass deposition rates ($L_{\rm{NSC}}$ and $\dot M_{\rm{NSC}}$) relate to  the wind adiabatic terminal speed ($V_{\rm{A},\infty}$) by $L_{\rm{NSC}}=0.5\dot M_{\rm{NSC}}V_{\rm{A},\infty}^{2}~\sim 3\times 10^{40} (M_{\rm{NSC}}/10^{6}M_{\odot})$. The second expression arises when one scales the average results from Starburst99 \citep[SB99,][]{1999ApJS..123....3L}.  The wind adiabatic terminal speed is   
 an input parameter in the model, we assume in all cases that it is constant and equal to 1000 km s$^{-1}$. Note that this value is 2.5 times lower than the average adiabatic wind terminal speed \citep[see][]{2011ApJ...740...75W} for an instantaneous starbursts during the first 40 Myr.  Therefore, $V_{\rm{A},\infty}=1000$ km s$^{-1}$ implicitly means additional mass loading to the flow, which may be due to evaporation and destruction of preexisting high density molecular clouds and filaments, and/or evaporation of circumstellar discs forming low mass stars, see for example \citet{2003MNRAS.339..280S,2004A&A...424..817M}. We assume that mass loading is five times that inserted  by massive stars.

\section{The numerical approach}\label{s3}

The numerical models presented here are based on the finite-difference Eulerian hydrodynamic code ZEUS-3D version 3.5 \citep{1992ApJS...80..753S,2010ApJS..187..119C}.  All calculations were performed in spherical coordinates in 2D, with symmetry along the $\phi$- direction.   
 The set of hydrodynamic equations is
\begin{equation}
\frac{\partial\rho}{\partial t} + \nabla\cdot(\rho {\bf u})=q_{m},
\label{e21}
\end{equation}
\begin{equation}
\frac{\partial{\bf u}}{\partial t} + ({\bf u\cdot \nabla}){\bf u} =
-\frac{1}{\rho}{\bf \nabla}P-{\bf \nabla}\Phi + {\bf g}_{\rm rad},
\label{e22}
\end{equation}
\begin{equation}
\frac{\partial \epsilon}{\partial t} + {\bf \nabla}\cdot[{\bf u}(\epsilon + P + \rho \Phi)]=q_{e}-Q+H_{\rm AGN},
\label{e23}
\end{equation}

\noindent  and it is closed by the equation of state $P=(\gamma-1)e$, where $e$ is the internal energy density and $\gamma=5/3$ the adiabatic index. The total internal energy is $\epsilon=\rho u^{2}/2 + e$.  The mass and energy deposition rates per unit volume are $q_{m}=3\dot M_{\rm{NSC}}/(4\pi R_{\rm{NSC}}^{3} (1 + \eta_{\rm ml}))$ and $q_{e}=3L_{\rm{NSC}}/(4\pi R_{\rm{NSC}}^{3})$, respectively. The parameter $\eta_{\rm ml}$ represents mass loading. A small value of $V_{A, \infty}$ means that the mass in winds of individual stars is loaded by additional mass from the parental cluster.
The magnitude of the local acceleration due to gravity is $\mid\nabla \Phi\mid \equiv g_{\rm grav}=-GM(r)/r^{2}$, where $M(r)=M_{\rm{SMBH}}+M_{\rm{NSC}}(r/R_{\rm{NSC}})^{3}$ is the mass enclosed within a sphere of radius $r$. 
The terms ${\bf g}_{\rm rad}$ and $H_{\rm AGN}$  in equations (\ref{e22}) and (\ref{e23}), respectively, represent the radial acceleration due to radiation pressure and the heating rate per unit volume due to the X-rays from the central source,  the method used to calculate these terms is described in  Appendix \ref{A}. 
The cooling rate per unit volume is  $Q=n^{2}\Lambda(T,Z)$, where $n=\rho/\mu(T)m_{\rm H}$  is the gas number density,  $m_{\rm H}$ is the  proton mass, and $\Lambda(T,Z)$ is the cooling function,  which depends on temperature $T$ and metallicity $Z$.   In all cases solar metallicity is assumed. To compute $n$ we use an approximate treatment for the ionization degree: we take the values $\mu(T \ge10^{4})=14/23$ and $\mu(T<10^{4})=14/11$ as the mean mass per particle for ionized and neutral gas respectively. 

The model accounts for extremely fast cooling in all the computational domain and is considered in the calculation of the time step.  The cooling rate is computed with an updated routine of W08 that uses, for temperatures above the ionization temperature of hydrogen, the Raymond \& Cox cooling function tabulated by \citet{1995MNRAS.275..143P}, and for T$\lesssim 10^{4}$ K the \citet{2002ApJ...564L..97K} cooling function: $\Lambda(T)=\Gamma_{\rm KI}~[10^{7}\exp(-1.184\times10^{5}/(T+10^{3}))+1.4\times10^{-2}T^{1/2}\exp(-92/T)]$ erg cm$^{3}$ s$^{-1}$, where $\Gamma_{\rm KI}=2.0\times10^{-26}$ erg s$^{-1}$. The minimum allowed temperature in the simulations is assumed to be T=100 K.   Note that according to \citet{2006ApJ...653.1266J}  the gas is thermally stable at temperatures where the slope of the cooling curve in the space $\log(\Lambda(T))~vs~T $ is $\ge 1$, what led  \citet{2009MNRAS.393..759S} to identify  five stable zones in the Plewa (1995) cooling function.

The flow was modeled following the prescription given in W08, which explicitly considers a continuous replenishment of mass and internal energy in all cells within the starburst volume at rates $q_{m}$ and $q_{e}$, respectively.
The  inserted mass is subject to the gravity force of the SMBH as well as from the NSC, like in HZ10, however here the mass is injected with an angular momentum that  corresponds  to the solid body rotation of the cluster.
 In summary, the procedure applied to each cell within the cluster volume at every time step is:

%--------------------------------------------------------------
\begin{enumerate}

\item{The radial velocity of the flow $v_{r}$ is updated according to $v_{r}=v_{r}+(g_{\rm grav}+g_{\rm rad})dt$. where $g_{\rm grav}$ and $g_{\rm rad}$ are the local acceleration due to gravity and radiation pressure.}

\item{The density and  total energy in a given cell are saved to $\rho_{\rm{old}}$ and $e_{\rm{tot, old}}$.}

\item{The mass is inserted so that  $\rho_{\rm{new}}=\rho_{\rm{old}}+\delta\rho$, where $\delta\rho=(1+A_{\rm{noise}}\zeta)q_{m}dt$ is the injected mass per unit volume. The  mass $\delta\rho $ is inserted  with rotation velocity $v_{\rm{rot}}$, along the $\phi-$ direction, assuming solid-body rotation for the star cluster:  $v_{\rm{rot}}=\omega r \sin \theta$, where $\omega$ is the angular frequency, $r$ is the distance from the center,  and $\theta$ the angle from the polar axis.}

\item{The velocity is corrected so that the momentum is conserved:
 ${\bf v}_{\rm{new}}={\bf v}_{\rm{old}}\rho_{\rm{old}}/\rho_{\rm{new}}+\hat{\phi} v_{\rm{rot}}\delta\rho/\rho_{\rm{new}}$, where the components of the velocity vector {\bf v} are $v_{r}, v_{\theta}$ and $v_{\phi}$.  $\hat{\phi}$ is the unit vector in the direction $\phi$.}

\item{The internal energy is corrected to conserve the total energy $e_{i, \rm{mid}} = e_{\rm{tot, old}}-\rho_{\rm{new}}{\bf v}_{\rm{new}}^{2}/2$.} 
\item{The new energy is inserted in a form of internal energy $e_{i,\rm{new}}= e_{i, \rm{mid}} +(1+A_{\rm{noise}}\zeta)q_{e}dt $.}
\item{The AGN-heating $H_{\rm AGN}$ is included by increasing the internal energy in each grid cell by $H_{\rm AGN}dt$.} 
\end{enumerate}
\noindent In steps 3 and 6, $\zeta$ is a random number from the interval (-1,1) generated each time step it is used, and $A_{\rm{noise}}$ is the relative amplitude of the noise. The inclusion of the noise is necessary to break the spherical symmetry imposed by the initial conditions, an analysis of the effects of such noise is presented in W08 in the case of star cluster winds. Here we used their  recommended value $A_{\rm{noise}}=0.1$ in all our simulations.  

%-------------------------------------------------------------- 
\subsection{Initial and boundary conditions}\label{s31}
 After an initial relaxation period, the solution reaches a  steady state with a quasi-stationary wind blowing from the outer parts of the cluster and mass accumulation at an approximately constant rate in the inner region. To make the transition as short as possible, the models start from the stationary adiabatic wind solution of a star cluster with mass $M_{\rm NSC}=10^{6}M_{\odot}$, adiabatic wind terminal speed $V_{\rm{A},\infty}=1000$ km s$^{-1}$, and with radius in each case as given in Table \ref{t1}.  
The boundary conditions are set open at both $r$-boundaries and reflecting at both $\theta$-boundaries. We used the scaled grid option in $r$ and a uniform one in $\theta$. The computational domain and the number of zones in each direction were selected such that $\Delta r \sim r\Delta \theta$, which conserves the shape of the zones and provides a higher resolution closer to the center.  

%----------------------------------------------------------------------------------------------------
\begin{table}[!h]
\tabletypesize{\tiny}
\begin{center}
\caption{Selected models and results from the simulations. \label{t1}}
\begin{tabular}{ccccccccccc}
\tableline\tableline
No. & $M_{\rm{NSC}}$ & $M_{\rm{SMBH}}$ & $R_{\rm{NSC}}$ & $v_{\rm{rot}}$ & Rad.& $R_{\rm{st}}$ & $R_{\rm T,num}$& $R_{\rm T,anl}$  & $\chi_{24}$&$\chi_{\rm cold,22}$  \\
\cline{7-9}\\
    & ($M_{\odot}$) & ($10^{6}M_{\odot}$) & (pc) & (km~s$^{-1}$)& &\multicolumn{3}{c}{(pc)}&  &\\
\tableline
1   & $3.3\times 10^8$        & 1               & 10   &50 &   N         & 9.2& $\sim$2\phn& 1.95    & 0.76\phn &0.92\\  
2   & $3.3\times 10^8$        & 1               & 10   &50 &   Y         & 9.2& $\sim$2\phn& 1.95    & 0.93\phn &0.86\\  
3   & $3.3\times 10^7$        & 1               & 10   &50 &   Y          & $\cdots$\phn& $\cdots$\phn& $\cdots$        & $\cdots$ &$\cdots$\\ 
4   & $3.3\times 10^8$        & 1               & 40   &50 &   Y         &31.3 & 10.5& 9.7  & 0.71 &0\\ 
5   & $3.3\times 10^8$        & 10\phn      & 40   &50 &   Y         & $\cdots$\phn& $\cdots$\phn& $\cdots$   & $\cdots$\phn &$\cdots$\\ 
\tableline
\end{tabular}
\tablecomments{Summary of the models. The first column marks the label of each model, where model 2 is our reference model; $M_{\rm NSC }$ and $M_{\rm SMBH}$ are the masses of the NSC and the SMBH respectively; $R_{\rm NSC}$ is the radius of the star cluster; $v_{\rm rot}$ is the maximum rotation velocity at the star cluster surface; the labels Y  and N, indicate whether a model includes the radiation from the central source or not.  $R_{\rm{st}}$ is the  $\theta-$averaged value of the  stagnation radius measured in the simulations at a time when the solution is steady; $R_{T}$ is the radius of the centrifugal barrier, where the subscripts $num$ and $anl$ denote the results from the numerical simulations and the analytic prediction (Eq. \ref{a52}) respectively,  numerically it corresponds to the time averaged radius of the point of maximum density; $\chi_{24}$ indicates the time averaged of the fraction of the sky covered by column densities larger than $10^{24}$ cm$^{-2}$, and $\chi_{\rm cold,22}$ gives the fraction of the sky covered only by cold gas ($T<$ 1500 K) with column density larger than $10^{22}$ cm$^{-2}$. }
\end{center}
\end{table}
%----------------------------------------------------------------------------------------------------

\section{Results}\label{s4}

We have calculated a set of models (see Table \ref{t1}) with  NSC and 
SMBH parameters in the range given by the observations (see for example 
Figure 19 in \citet{2010ApJ...714..713S}). Models  1 and 2 have a star 
cluster radius $R_{\rm{NSC}}$= 10 pc and mass  $M_{\rm{NSC}} = 
3.3\times10^{8}M_{\odot}$, respectively, both with a $10^{6}M_{\odot}$ SMBH. In 
model 1, the radiation from the central source is not considered. Hereafter we 
refer to model 2 as the reference model.  In model 3 a star cluster is 10 
times less massive than in the first two cases. Models 4 and 5 consider more 
extended clusters with $R_{\rm{NSC}}$= 40 pc, $M_{\rm{NSC}} = 
3.3\times10^{8}M_{\odot}$ and SMBHs of $M_{\rm{SMBH}}=10^{6}M_{\odot}$ and 
$M_{\rm{SMBH}}=10^{7}M_{\odot}$, respectively. The computational domain extends 
radially from 0.5 pc to 15 pc in the case of models 1 -- 3, and from 1 pc to 
60 pc for models 4 and 5. The axial extent goes from $\pi /2 -1.3$ to 
$\pi /2 +1.3$ radians in all calculations. The complete set of relevant 
parameters for the simulated models are given in Table \ref{t1}.

\subsection{The hydrodynamic solution}\label{s41}
\subsubsection{Case without a central source of radiation (model 1)} 
We start with a description of  model 1, in which the radiation from the central source is not considered.   
According to TT07 and W08 such NSC evolves in a catastrophic cooling regime. Here it is shown that the inclusion of gravity of the NSC+SMBH and the angular momentum of the inserted matter lead to the formation of a filamentary/clumpy torus. 

\begin{figure}[!htb]
\centering
\includegraphics[scale=0.7]{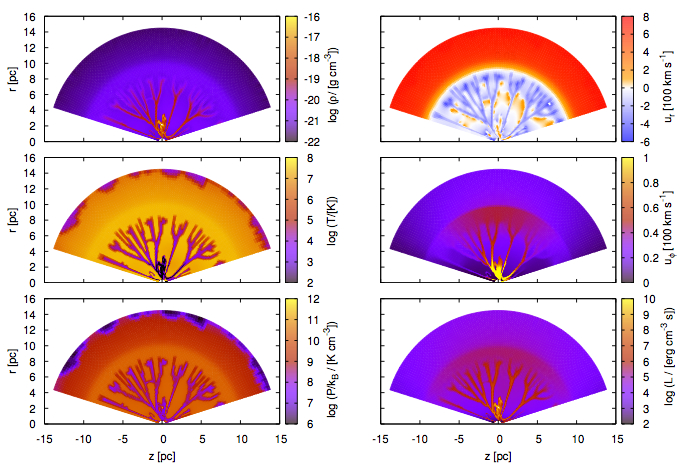}
\caption{Filamentary torus resulting from model 1, snapshots  at 1 Myr. The torus is composed by a collection of cold filaments and a dense core located at the centrifugal barrier, at 2 pc. In the left column, from top to bottom, the plotted quantities are logarithms of density, temperature and pressure divided by the Boltzmann constant $k_{B}$. In the right column, from top to bottom, the panels corresponds to radial velocity, tangential velocity, and logarithm of angular momentum.}
\label{f1}
\end{figure}

Initially, the average gas density in the central part grows rapidly due to the 
mass deposition   by the stellar cluster. Radiative cooling is enhanced 
because of its squared dependence on density and thus temperature in the densest 
zones drops. As soon as temperature  decreases to approximately $10^{6}$ K, 
{\it free-bound} and {\it bound-bound} transitions become the major cooling 
processes. Consequently, when the temperature in some region approaches 
$\sim 3\times 10^{5}$ K the cooling rate increases steeply and the thermal 
instability starts to operate. This lowers the temperature to 10$^2$ K, and 
 decreases the pressure by three orders of magnitude from that 
($P\sim 3\times10^{-6}$ dyne cm$^{-2}$) in the hot ($\sim10^{7}$ K) gas. The 
cold regions are compressed by the surrounding hot gas into dense filaments/clumps. 
After an initial relaxation period, the stagnation radius $R_{\rm st}$,  
which is defined by radiative cooling, remains almost constant. Note that in 
our model this radius determines the amount of gas that flows inwards and that 
later accumulates close to the center and that $R_{\rm st}$ is different 
from the Bondi radius which is defined by the mass of the central black hole and 
by the sound speed in the surrounding ISM.

Figure \ref{f1} presents frames of the distribution of the  hydrodynamical 
variables in the whole computational domain. The density distribution presents 
mainly  two-phases: hot (few 10$^{7}$ K) gas with low density, and  cold 
(T=10$^{2}$ K) gas in the densest  zones in a form of  filaments and clumps.    
The thermal pressure gradient within the stagnation volume ($R_{\rm st}=9.2$ pc) 
is  not high  enough to push the cold gas out from the cluster volume and 
instead the cold parcels of gas begin to stream toward the center because of the 
force of gravity. Due to angular momentum conservation, the rotation velocity of 
the inflowing gas increases as it approaches the center, to about 200 km s$^{-1}$
at $\sim 2$ pc. Such fast rotation prevents the gas to flow further inwards and  
favors the accumulation of mass around the centrifugal barrier. We identify the 
collection of cold filaments and the dense core at the centrifugal barrier as the
torus responsible for the obscuration of the central SMBH. While all this 
happens in the central part of the cluster, a stationary cluster wind reaching 
a terminal speed of about 800 km s$^{-1}$ is well sustained above $R_{\rm{st}}$ 
in all simulated cases. Such winds may prevent
the inflow of matter from larger scales onto the NSC. This suggests that 
NSC-SMBH may evolve in isolation when the  feedback from massive stars is highly 
active.

\subsubsection{Reference model (model 2)}      

\begin{figure}[!ht]
%\centering
\hspace{0cm}
\includegraphics[scale=0.8]{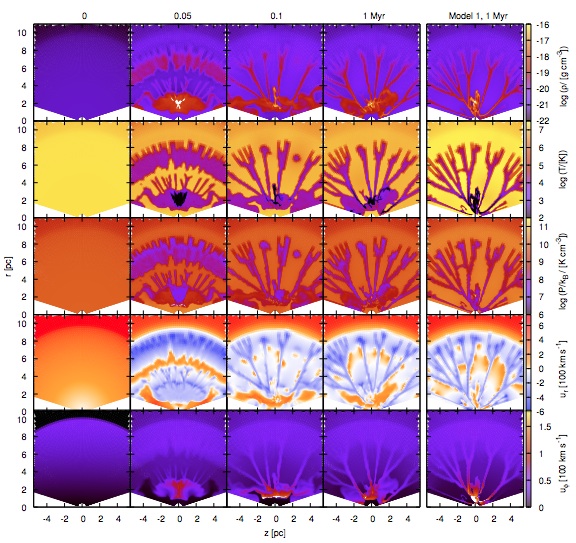}
\caption{Time sequence of the distribution of the hydrodynamic variables in the reference model. All panels show the $z-r$ plane, with the $z$ axis running horizontally. Each column displays frames at the time indicated on the top, for each variable indicated at the end of each row.  As a comparison, the last column shows the corresponding variables for model 1 at $t=1$ Myr.}
\label{f2}
\end{figure}

The  input parameters in this model are the same as in model 1 except that 
in this case we consider the effect of the X-ray radiation  from the 
central SMBH. Figure \ref{f2} presents  a sequence of frames of density, 
temperature, pressure, radial $u_{r}$ and tangential $u_{\phi}$ components of 
velocity for the reference model.  For comparison, the panels in the right  most 
column display the distribution of the corresponding variables at 1 Myr for the 
case without central source of radiation (model 1).  As one can note, the  
stagnation radius remains the same in both models. This implies that the same  
amount of mass is accumulated in the central part of the cluster. However, the 
inner zone is  affected by the central radiation field, which prevents the gas 
from the fast cooling at the very center ($r<$1 pc).  Therefore in the
reference model the gas there remains hot (at $\sim 10^{7}$ K, see the second 
row in Figure \ref{f2}). However, the  flow at larger radii (1 pc $<r<$ 3 
pc) is still dominated by cooling and a substantial amount of cold  gas is 
accumulated there forming a filamentary/clumpy torus  with a dense core 
supported by rotation at the centrifugal barrier. A fraction of this gas is 
ionized by the SMBH radiation, so, it remains warm (1500 K $<T<10^{5}$K). 
Note that the main difference between models 1 and 2 is due to the thermal 
pressure of the ionized gas in the central zone but not due to the radiation 
pressure. 

\begin{figure}[!ht]
%\centering
\hspace{2cm}
\includegraphics[scale=0.8,angle=0]{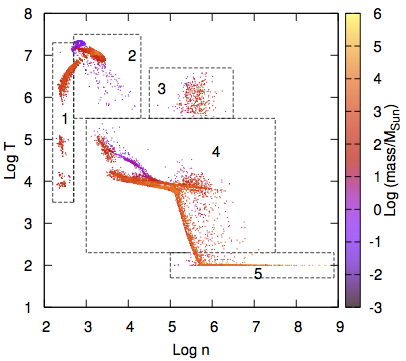}
\caption{Phase diagram for the standard model. Temperature against number density at $t=1$ Myr. The diagram was gridded into $70^{2}$ cells with the masses of the points depicted in color. We identify five components of the flow: 1- wind, 2- hot thermalized gas, 3- heated gas in the torus, 4- filaments and clumps, and 5- cold dense core of the torus. The gas tends to settle at stable regions in the cooling curve, in particular at around $10^4$K, however the squared dependence of the cooling rate on density leads to lower temperatures.}
\label{f3}
\end{figure}

Figure \ref{f3} shows the multi-phase medium that results from the 
simulation in the reference model. We identify five regions in the $\log T$ vs 
$\log n$ plane which represent different components of the flow. Region 1 
corresponds to the wind, whose temperature drops due to the expansion and 
radiative cooling. One can also find in this region temperatures that coincide 
with two stable regions (around $10^4$ and $8\times 10^4$K) of the 
\citet{1995MNRAS.275..143P} cooling function noticed by 
\citet{2009MNRAS.393..759S}. Region 2 presents hot ($\sim 10^7$K) rarefied gas 
resulting from shock-shock collisions and radiative heating. Region 3 results 
from the radiative heating of the surfaces of dense clumps in the torus. This 
region does not exist in the case without central source of radiation. Region 4 
corresponds to the collection of filaments which tend to settle in one of the 
stable branches of the cooling function. Eventually these cool down to the 
minimum allowed temperature in the simulation due to the increase of density 
as gas moves towards the center. Some high density ($ n>10^5$ cm$^{-3}$ ) zones 
are heated up to $\sim 10^{4}$ K by the AGN radiation. These zones are also not 
present in model 1 which does not include the central source of radiation. 
Region 5 corresponds to the  core of the torus at the centrifugal barrier where 
most of the mass is concentrated. The majority of the gas is in the cold phase (T$<$1500 K; see Figure 
\ref{f4}).

\subsubsection{Other models}
In model 3, the stellar cluster is 10 times less massive than that in the 
reference model.  Therefore, the density of the inserted gas is an order of 
magnitude lower. In this case, the central source of radiation heats up all the 
gas within the cluster volume up to few  $10^{7}$ K and prevents the formation 
of thermally unstable zones (clumps).  Therefore the reinserted matter does
not form torus in this case.  

In model 4, the stellar cluster has the same mass as in the reference model, 
however the cluster is more extended: $R_{\rm NSC}=40$ pc.  Therefore, the 
density of the inserted mass is also lower compared to the reference model. 
Nevertheless, thermal instabilities occur in the densest regions  where the 
accumulated mass forms a torus. However, in this case the torus is composed only 
of warm gas.  On the other hand,  some clumps formed close to $R_{\rm st}$ 
eventually join the cluster wind and leave the cluster,  reducing the mass 
accumulation rate.

In model 5 we consider an extended cluster as in model 4 but, in this case the 
SMBH is 10 times more massive and therefore more energetic than in all previous 
cases. The strong radiation field keeps the matter within the whole cluster hot 
 what does not allow to form a torus, however a powerful wind as in model 3 is generated. Such cases resemble adiabatic calculations given the impact of radiation. 
    
\subsection{Column density and obscuration fraction} \label{s42}
In the AGN models it is supposed that a torus, uniform or clumpy, blocks the 
light coming from the accretion disc. 
The amount of obscuring gas is usually quantified by the column density, i.e. 
the number of particles per unit area along the line of sight $N=\int{ndl}$.  
The optical/UV radiation is strongly attenuated above  $N=10^{22}$ cm$^{-2}$. 
If $N> 10^{24} \: {\rm cm^{-2}}$, the opacity is high enough to block even X-ray 
photons, in such a case the AGN  
 is said to be Compton-thick \citep[and references therein]{2004ASSL..308..245C}.      
There  is observational evidence that suggests that a large fraction of 
AGNs in the local universe are obscured by Compton thick gas 
\citep{1998A&A...338..781M,1999ApJ...522..157R,2000A&A...355L..31M} and 
 that most of them are associated with Seyfert 2 galaxies 
\citep{1999ApJ...522..157R}.
   
\begin{figure}[!htbp]
\centering
\includegraphics[scale=1]{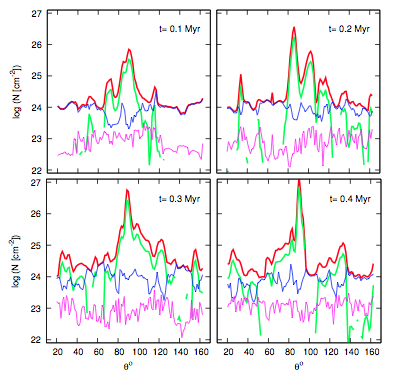}
\caption{Column densities as seen from the central SMBH for the reference model.  The red line displays the total column density along each line of sight. The magenta line shows the column density for hot gas, T $\ge 3\times 10^{5}$ K. The blue line represents the column density for warm phase, 1500 K $<$ T $<3\times 10^{5}$K. The column densities of cold matter (T $\le$ 1500 K) are shown by green lines. The cold gas does not cover all the sky.} 
\label{f4}
\end{figure}

Figure \ref{f4} displays for the reference model the column density at different evolutionary times as a function of the viewing angle as seen from the central SMBH. The line of sight along the equator corresponds to $\theta=90^{\circ}$. Different colors correspond to different gaseous components. Note that the column density of the warm gas (blue line) is high enough ($\sim 10^{24}$cm$^{-2}$) to block a large fraction of the X-ray radiation, and thus may turn the AGN Compton thick. The cold component (green line) presents gaps and high variability in its covering angle, which implies that UV photons can escape from the torus through the  holes
in the neutral gas and an observer can see eventually directly to the center. Such events offer a natural explanation to  the ``mutation" of optically classified Seyfert 2 to Seyfert 1, and vice versa. \citet{1994IAUS..159..438A} and \citet{1999ApJ...519L.123A} give examples of such kind of objects.

\begin{figure}[!htbp]
%\centering
\hspace{0cm}
\includegraphics[scale=0.6,angle=0]{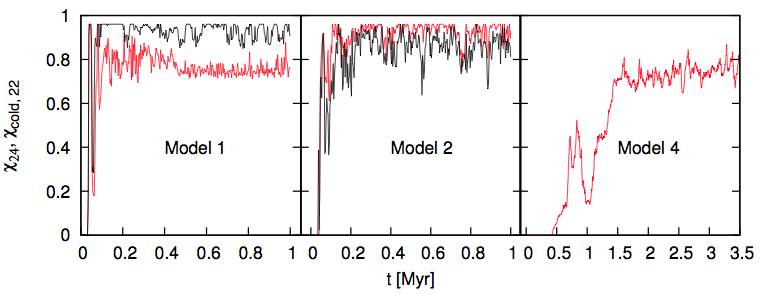}
\caption{Fraction of the sky covered by optically thick gas to optical/UV and  X-ray radiation as a function of time. Black lines represent the fraction of the sky,  $\chi_{\rm cold, 22}$, covered by  cold gas in the torus with column densities $N \ge 10^{22}$ cm$^{-2}$. Red lines shows the fraction of the sky, $\chi_{\rm 24}$, covered by  total column densities $N \ge 10^{24}$ cm$^{-2}$.} 
\label{f5}
\end{figure}

 Figure \ref{f5} presents the fraction of the sky covered by two different 
column densities as a function of time for models 1, 2 and 4: $\chi_{24}$ (red lines) represents  the fraction of the sky covered  by  total column densities $N \ge 10^{24}$ cm$^{-2}$,   $\chi_{\rm cold, 22}$ (black lines) shows the fraction of the sky covered by cold gas with column density $N \ge 10^{22}$ cm$^{-2}$. In the case of model 1,  the time average values are $\chi_{\rm cold, 22}\sim 0.93$ and $\chi_{24}\sim 0.76$. In the reference model, $\chi_{\rm cold, 22}\sim 0.86$ and $\chi_{24}\sim 0.92$. Note that $\chi_{24}$ is larger in the reference model compared to model 1. The same tendency was found by \citet{2012ApJ...758...66W} for Eddington ratios in the range 0.01-0.1, where more luminous AGNs are obscured over larger solid angles.  The fraction $\chi_{\rm cold, 22}$ in the reference model is reduced due to ionization by the central source.
In the case of model 4 the torus is composed only of warm gas, with average $\chi_{24}\sim 0.75$ after 1.5 Myr.     

\subsection{Comparison with analytic predictions}\label{s43}
The thermally unstable gas inserted within the stagnation radius is attracted by the gravity towards the cluster center. Due to its angular momentum it accumulates around the centrifugal barrier, where the rotation balance the gravitational attraction.
Here we give the analytic formula for such radius and show that it is in a good agreement with the numerical 
results. 

The radius where the mass accumulates, $R_{T}$, is determined by the angular momentum of the matter inserted within the stagnation radius $R_{\rm{st}}$ and by the central gravitational potential of the SMBH + NSC. Here we neglect the effect of the radiation pressure. The value of $R_{\rm st}$ is defined by  radiative cooling, which  depends on the mass and compactness of the cluster (see TT07 \& W08). 
Thus, in order to estimate the position of the centrifugal barrier we need to know $R_{\rm{st}}$ for a given rotation velocity of the star cluster.  In the following we derive an analytic relation for  $R_{T}$ by assuming a star cluster in solid-body rotation. 

Let us consider the total mass inserted within the stagnation zone at some 
instant. Then the specific angular momentum of a rotating parcel of gas 
is $j=\omega R^{2}$, where $R$ is the projection of $r$ on the equatorial 
plane, i.e., the distance to the rotation axis. An integration over the mass 
within the stagnation volume $V_{\rm{st}} = 4\pi R_{\rm{st}}^3/3$, gives the average 
specific angular momentum inserted within $R_{\rm{st}}$ at some instant:
\begin{equation}
j_{\rm{av}}=\frac{1}{V_{\rm{st}}}\int_{V_{\rm{st}}} {\omega R^{2}}dV.
\end{equation}    
\noindent
%-----------------------------------------------------------------------
Then by considering $\omega=$~constant and using spherical coordinates one
obtains:
%----------------------------------------------------------------------- 
\begin{equation}
j_{\rm{av}}=\frac{3\omega}{4\pi R_{\rm{st}}^{3}}\int_{V_{\rm{st}}}{r^{4}\sin^{3}(\theta) d\theta d\phi dr}=\frac{2}{5}\omega R_{\rm{st}}^{2}.
\end{equation}
%-----------------------------------------------------------------------
One expects the accumulation of mass at the centrifugal barrier, i.e., at the radius  $R_{T} = j_{\rm{av}} / v_{\rm K}$,
where $v_{\rm K}=(GM(R_{T})/R_{T})^{1/2}$ corresponds to the Keplerian velocity of
the gas orbiting the  mass  $M(R_{T}) = M_{\rm{SMBH}} + M_{\rm{NSC}}(R_T)$.
This leads to the algebraic equation:
%-----------------------------------------------------------------------
\begin{equation}
R_{T}^{4}+\frac{M_{\rm{SMBH}}}{M_{\rm{NSC}}}R_{\rm{NSC}}^{3} R_{T}-\frac{j_{\rm{av}}^{2}R_{\rm{NSC}}^{3}}{GM_{\rm{NSC}}}=0.
\label{a51}
\end{equation}
%----------------------------------------------------------------------- 
\noindent The physical solution of equation (\ref{a51}) is:
%----------------------------------------------------------------------- 
\begin{equation}
R_{T}=- \left(\frac{p}{2}\right)^{1/2} + \left[-\frac{p}{2}+(p^{2}+b)^{1/2}\right]^{1/2},
\label{a52}
\end{equation}
%----------------------------------------------------------------------- 
\noindent where

\begin{equation}
p=\left[ \frac{a}{2} + \left(\frac{a^{2}}{4}+\frac{b^{3}}{27}\right)^{1/2}\right]^{1/3}
-\left[-\frac{a}{2} + \left(\frac{a^{2}}{4}+\frac{b^{3}}{27}\right)^{1/2}\right ]^{1/3},
\label{a53}
\end{equation}

\noindent with 
\begin{equation}
a= \frac{M_{\rm{SMBH}}^{2}}{8M_{\rm{NSC}}^{2}}R_{\rm{NSC}}^{6}, \qquad
b= \frac{j_{\rm{av}}^{2}R_{\rm{NSC}}^{3}}{GM_{\rm{NSC}}}.
\end{equation}

Thus,  $R_{\rm{st}}$ is the key parameter to estimate the centrifugal barrier, 
because it determines the amount of mass and angular momentum inserted within 
the stagnation zone for a given  set of parameters: $R_{\rm {NSC}}$, $M_{\rm{NSC}}$, 
$M_{\rm{SMBH}}$, and $\omega$. In our calculations, $R_{\rm{st}}$ is self-consistently 
determined.  It splits the cluster into two zones: the inner one where the 
reinserted matter is accumulated and the outer one where the star cluster wind is
formed. It is worth to note that $R_{\rm st}$ is almost independent on the SMBH 
mass and may have a non-zero value even if $M_{\rm SMBH} = 0$.
This makes our 
approach different from all the modified Bondi models used so far to estimate 
the size of accretion discs or torus, see for example \citet{1976ApJ...210..377U,2003ApJ...582...69P,2005ApJ...618..757K,2010AstL...36..835I},  where the solution depends on the size of the so-called Bondi radius which is a function  of the SMBH mass and the sound speed in the surrounding ISM. 

\begin{figure}[!ht]
\centering
\includegraphics[angle=0,scale=0.8]{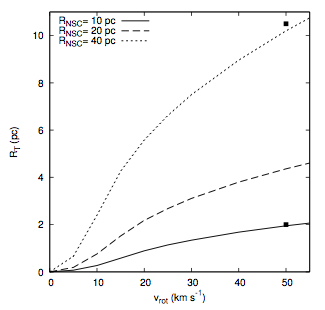}
\caption{Analytic prediction for the centrifugal barrier. The radius $R_{T}$  where mass accumulates as a function of the maximum rotation velocity of star clusters of mass $M_{\rm{NSC}}=3.3\times 10^{8}M_{\odot}$ and central SMBH of  mass $M_{\rm{SMBH}}=10^{6}M_{\odot}$, for different star cluster radius. The squares represent the reference model (model 2) and model 4, see Table \ref{t1}.}  
\label{f6}
\end{figure}

Figure \ref{f6} shows the analytic predictions for the centrifugal barrier as 
a function of the maximum rotation velocity  of the NSC, i.e., the rotation 
velocity at $r=R_{\rm{NSC}}$ and $\theta=90^{\circ}$. Different lines correspond to NSCs of 
different radii, all of them with a central SMBH of $10^{6}~M_{\odot}$. The squares represent the average values in the case of the reference model ($R_{\rm NSC}$ = 10 pc) and model 4 ($R_{\rm NSC}$ = 40 pc).  In all 
cases the mass of the NSC corresponds to that in the reference model 
($M_{\rm{NSC}}=3.3\times 10^{8}M_{\odot}$). As one can  expect  $R_T$ 
increases monotonically with the assumed rotation velocity of the cluster. 
Note that in the range of parameters here used, if one considers more 
extended clusters at a fixed $v_{\rm{rot}}$, the absolute value of the stagnation radius is larger,  but the ratio $R_{\rm{st}}/R_{\rm{NSC}}$ is smaller. Therefore, the specific angular momentum inserted is higher and the mass accumulates at a larger distance from the center.     
If the NSC is very extended, the impact of cooling is less important  and even the absolute value of $R_{\rm{st}}$ gets smaller, and eventually the NSC could be in a quasi-adiabatic regime\footnote{See,  
\citet{2004ApJ...610..226S} and  \citet{2007ApJ...658.1196T} for a  
discussion on the threshold energy which separates star clusters evolving in the catastrophic cooling regime from those evolving in a quasi-adiabatic regime.}. In such  cases  $R_{\rm{st}}$ tends to a very small value (S04, TT07, W08) and is mainly defined by the gravitational potential (S08 and HZ10). 

%----------------------------------------------------------------------------------------------------
\begin{table}[!htb]
\tabletypesize{\tiny}
\begin{center}
\caption{Mass accumulation rate. \label{t2}}
\begin{tabular}{ccccccccccc}
\tableline\tableline
Model & $M_{\rm{NSC}}$ & $M_{\rm{SMBH}}$ & $R_{\rm{NSC}}$ & $R_{\rm{st}}$& & $\dot M_{\rm{NSC}}$ & $\dot M_{\rm{wind}}$ & $\dot M_{\rm{acc}}$& $\dot M_{\rm{in}}$& \\
\cline{4-5} \cline{7-10} \\
    & ($M_{\odot}$) & ($10^{6}M_{\odot}$) & \multicolumn{2}{c}{(pc)}& &\multicolumn{5}{c}{($M_{\odot}$~yr$^{-1}$)}  \\
\tableline
1   & $3.3\times 10^8$        & 1               & 10   & 9.2& &31.2 & 7.3    & 22.1 &1.8 &\\ 
2   & $3.3\times 10^8$        & 1               & 10   & 9.2& &31.2 & 7.2    & 23.9 & 0 &\\  
4   & $3.3\times 10^8$        & 1               & 40   & 31.3& &31.2 & 15.9   &14.6  & 0&\\ 
\tableline
\end{tabular}
\tablecomments{Mass flow rates for models with typical parameters of NSC and SMBH in Seyfert type galaxies. A substantial amount of the total inserted mass ($\dot M_{\rm NSC}$) remains locked within the stagnation radius ($R_{\rm st}$), and accumulates at a rate $\sim \dot M_{\rm{acc}}$ in the torus. Note that a good fraction of the total injected mass blows out in the NSC wind at rate $\dot M_{\rm{wind}}$. The inflow rate through the inner boundary of the computational domain is denoted by $M_{\rm in}$. }
\end{center}
\end{table}
%----------------------------------------------------------------------------------------------------

\subsection{Mass accumulation rate} \label{s44}

In all simulations the hydrodynamic solution reaches a steady state. In the case 
of our reference model it happens after $\sim$0.1 Myr and at about four times 
longer  time in the case of 40 pc clusters. From then onwards the matter 
inserted in the region $R_{\rm st}< r < R_{\rm NSC}$ flows  through $R_{\rm NSC}$
and leaves the cluster as a stationary wind  which stops the income of 
matter from a large scale in the galaxy. On the other hand, the mass that remains
locked within the stagnation volume streams toward the center and accumulates 
around the centrifugal barrier  practically at a constant rate. Table \ref{t2} 
presents the corresponding rates.

\begin{figure}[!htbp]
%\centering
\hspace{0cm}
\includegraphics[angle=0,scale=0.6]{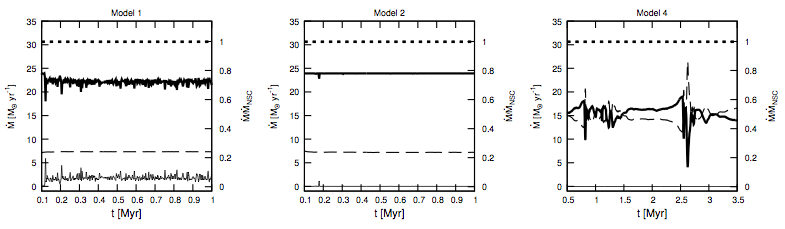}
\caption{Mass deposition, accumulation and outflow rates.  Left and right axis show scales of absolute  and relative quantities, respectively. Dotted  lines represent $\dot M_{\rm NSC}$, thick solid lines display $\dot M_{\rm acc}$,  dashed lines shows $\dot M_{\rm wind}$, and thin solid lines represent $\dot M_{\rm in}$. The average values are given in Table \ref{t2}.}
\label{f7}
\end{figure}

Figure \ref{f7}  presents for models in Table \ref{t2}, the absolute and relative (normalized to the star cluster mass input rate)
values  of the rates of mass deposition within the volume of the cluster
($\dot M_{\rm{NSC}}$), mass accumulation in the torus ($\dot M_{\rm{acc}}$), 
 mass carried away by the wind ($\dot M_{\rm{wind}}$), and mass inflow  towards 
the center through the inner zone of the computational domain ($\dot M_{\rm{in}}$). 
 In Model 1:  $\sim 71\%$ of the total inserted mass accumulates in the torus, 
$\sim 23\%$ leaves the cluster as a wind. About 6\% of the inserted  mass escape 
through the inner zone of the computational domain.   In the reference model 
(model 2): $\sim$ 77\% of the total inserted mass goes into the torus and $\sim$ 
23 \% goes into the cluster wind. Model 4: approximately one half of the inserted 
mass leaves the cluster as a wind, the rest accumulates in the torus.   In this 
case some clumps escape from the cluster, producing peaks in  $\dot M_{\rm{wind}}$ 
(dashed line) with the corresponding response in $\dot M_{\rm{acc}}$ (solid line).
Note that in models 2 and 4,  radiation pressure prevents the inflow of mass through 
the inner boundary.  The actual value of the accretion rate onto  the SMBH is 
beyond the scope of this paper as  the inflow of gas to the central black hole is 
inhibited by angular momentum when the viscous processes are negligible, 
\citet{2012arXiv1206.4059L}. Note that in all cases the rate of mass accumulation is 
given by  $R_{\rm st}$.   

The mass of the torus  at a given time can be estimated from the average 
$\dot M_{\rm acc}$. For example, at 1 Myr it reaches $2.39\times10^{7}M_{\odot}$ in 
the reference model, and $1.46 \times 10^{7} M_{\odot}$ in  the case of model 4 for 
the same period.  Due to the additional mass loading considered in our models, the 
mass of the torus grows substantially. Such torus is gravitationally unstable. An 
estimate of the Toomre parameter, $\texttt{Q}\equiv \Omega c_s/3G\Sigma$,  for 
the obscuring structure in our reference model in the region centered at the 
centrifugal barrier ($R_T\simeq 2$ pc) with $\Delta R = 0.5$~pc, sound speed 
$c_s\simeq$1 km s$^{-1}$, Keplerian rotation at frequency $\Omega \sim 1.4 \times 
10^{-12}$ s$^{-1}$ and surface density $\Sigma = 140$ g cm$^{-2}$ results in 
$\texttt{Q}\sim 5\times 10^{-3}$. Therefore, the mass accumulation should lead to a 
continuous star formation, which may amplify the effect of the stellar feedback 
in the nuclear region of the host galaxy.

%------------------------------------------------------------------------------------
\section{Impact of the NSC wind on the host galaxy}
In all cases presented in Table \ref{t1}, the star cluster wind is sufficiently powerful as to significantly re-structure the host galaxy ISM leading perhaps to a thick ring along the plane of the galaxy, and to a super galactic wind along the host galaxy symmetry axis \citep[as in][]{1997ApJ...478..134T, 1998MNRAS.293..299T}.

A simple estimate of the wind power can be obtained from its ram pressure 
$P_{\rm ram}=\rho u^2$ at the starburst edge. This is in all cases many orders of 
magnitude larger than the typical ISM pressure in our Galaxy ($\sim 10^{-12}$ dyne 
cm$^{-2}$). For example, for models 1 and 2, $P_{\rm ram}\simeq 1.5\times 10^{-6}$ 
dyne cm$^{-2}$; and about half this value in  model 3.  In models 4 and 5, 
$P_{\rm ram}=$  1.4 and 2.5$\times 10^{-7}$ dyne cm$^{-2}$, respectively.  Such values 
are comparable with the  ram pressure of the freely falling gas, $P_{\rm ff}=\dot 
M_{\rm in} u_{\rm ff} /(4\pi R_{NSC}^2)$ where $u_{\rm ff}= [2G(M_{\rm SMBH} + 
M_{\rm NSC})/R_{\rm NSC}]^{1/2}$ is the free fall velocity, in the torus formation
model of \citet{2012MNRAS.420..320H} with the inflow rate $\dot M_{\rm in} = 
10~M_{\odot}$ yr$^{-1}$ ($P_{\rm ff}\simeq 2.8\times 10^{-6}$ dyne cm$^{-2}$ for models 
with 10 pc radius and  $P_{\rm ff}\simeq 8.8\times 10^{-8}$ dyne cm$^{-2}$ for models 
with $R_{\rm NSC}=$ 40 pc). Thus the NSC winds have to build up super bubbles and 
probably super galactic winds preventing, in most cases, the falling of the ISM onto 
the NSC.  

\section{Conclusions}
We present 2D radiation hydrodynamic simulations of the gas reinserted by stars of a 
rotating young NSC with a central SMBH.  Our model considers explicitly the impact 
 which the stars from a NSC provide on the accretion flow. The model includes 
constant mass and energy deposition from stars  and assumes that the mass is 
inserted with non zero angular momentum. It takes into consideration  gravity from 
the central SMBH and from the NSC, and accounts for radiative cooling and the heating 
from a central isotropic source of X-ray radiation.

 Here we have shown that the combined effect of gravity from the SMBH+NSC 
and the angular momentum of the inserted mass results in the formation of a  dense 
structure (torus) inside the NSC, well within the stagnation radius $R_{\rm st}$, 
defined by radiative cooling. The torus  is only a few parsecs across, 
filamentary/clumpy, with a core at the centrifugal barrier. It is composed of gas in 
two phases: a cold phase (T$\le 1500$ K), where dust can survive as close as a couple of 
parsecs from the SMBH,  and a warm phase (1500 K$<$T$\le 3\times 10^{5}$K), 
maintained at this temperature by  heating from the central source of radiation. The 
torus is Compton thick and covers a large fraction of the sky, more than 80\% in our 
reference model. This obscuring structure is embedded into a low density hot gas.  

 Note that models developed by \citet{2002ApJ...566L..21W} and \citet{2009MNRAS.393..759S} 
and that are here discussed, lead to the flow of the cold gas towards
the central zone of the host galaxy. The inflow of molecular gas in the inner
$\sim 150$ pc of the Seyfert galaxy NGC 4051 was detected by
\citet{2008MNRAS.385.1129R} who suggested that the inflow occurs due
to the spiral arms that reach the nucleus of the galaxy.  
\citet{2012arXiv1210.2397G} presented the evidence for the 
inflowing gas in the broad line regions of four AGNs: Mrk 335, Mrk 1501,
3C 120 and PG2130+099. \citet{2012MNRAS.426.2859S} showed that the parsec-scale 
inflows do not result in significant absorption features in the X-ray spectra
since the ionization degree of the infalling gas is high. Therefore, the lack of such observations does not rule out our model.

The accreting mass accumulates in the central region almost at a constant rate, 
resulting after some time in a very massive torus.  As soon as it becomes 
gravitationally unstable a second generation of stars may form there leading to the 
formation of a stellar torus,  and thus matter would be continuously reprocessed into 
stars at a rate dictated by the mass accumulation. 

A necessary, but not sufficient, condition for the formation of the torus is that the 
 matter reinserted within the NSC evolves in a thermally unstable regime. However, 
the formation of the torus may be prevented by the strong central source of radiation
as it is the case in our models 3 and 5. 

In all cases  a powerful cluster wind is established outside the stagnation radius. 
Such winds can inhibit the income of gas from larger scale in the galaxy. This suggests
that during the starburst phase, when massive stars dominate the NSC feedback to the
host galaxy ISM, the NSC-SMBH interplay occurs in isolation.

\acknowledgments
 The authors express their thanks to the anonymous referee, whose constructive comments and proposal have helped improve this paper. This study has been supported by the Czech Science Foundation grant 209/12/1795 and by the project RVO: 67985815; the Academy of Sciences of  the Czech Republic and  CONACYT-M\'exico research collaboration under the project 17048: Violent star formation;   the CONACYT - M\'exico, research grants 131913 and 167169.   
H-Z. F. wishes to express his thanks to CONACYT-M\'exico  for additional support  through grants 162184 and 186720.

\appendix{}
\section{The heating rate and acceleration due to the central radiation field}\label{A}
We consider only the high frequency band of an AGN spectrum, which allows to take the advantage of a parametric form of the Compton ($\Gamma_{\rm Compton}$) and X-ray ($\Gamma_{X}$) heating functions given by \citet{1994ApJ...435..756B}.
By means of  a ray tracing we calculate the optical depth $\tau=\int^{r}_{r_{\rm in}}\kappa \rho dr$ at each radius $r$, where $r_{\rm in}$ is the inner boundary of the computational domain, $\rho$ is the local gas density, and $dr$ is the radial length of a  grid cell. We assume that the attenuation in the ionized gas is dominated by Thompson scattering, i.e. the opacity is $\kappa\simeq 0.4$ cm$^{2}$ g$^{-1}$,  and  two orders of magnitude higher \citep[see][]{2000ApJ...543..686P} for  the neutral gas. 

The optical depth $\tau$ is used to compute the X-ray flux $F_{X}=L_{X}e^{-\tau}/(4\pi r^{2})$, which is required to calculate the heating rate and the acceleration due to the radiation pressure. The amount of energy emitted in X-rays depends on the SED  and  the total luminosity of the source. For the average AGN SED given in  \citet{1997ApJS..108..401K}, one gets that about 8\% of the total luminosity is emitted in X-rays. Here we take this fraction and assume that the total luminosity corresponds to the Eddington limit for a given SMBH, therefore, we use $L_{X}=0.08L_{\rm Edd}$.

Once $F_{X}$ is known, the local acceleration due to the radiation pressure in equation (\ref{e22}) is computed as $g_{\rm rad}= F_{X}\kappa/c$, where $c$ is the speed of light. Then the radial velocity of the flow is corrected by $g_{\rm rad}dt$.

The heating rate per unit volume  is  $H_{\rm AGN}=n^{2}(\Gamma_{\rm Compton}+\Gamma_{X})$, it is included in equation (\ref{e23}) by increasing the internal energy in each grid cell by $H_{\rm AGN} dt$. Explicitly, $\Gamma_{\rm Compton}=8.9\times10^{-36}\xi(T_{X}-4T)$ and $\Gamma_{X}=1.5\times10^{-21}\xi^{1/4}T^{-1/2}(1-T/T_{X})$, both in units of erg s$^{-1}$ cm$^{3}$. Such functions depend on the temperature $T$ of the gas, the characteristic temperature $T_{X}$ of 10 keV X-ray radiation, and on the  ionization parameter  $\xi=4\pi F_{X}/n$ \citep{1969ApJ...156..943T}. The last parameter characterizes states of ionization equilibrium and depends on the local flux and the number density of particles within a grid cell. 

\bibliography{fhz}

\end{document}